\documentclass[intlimits,twoside,a4paper]{article}

\usepackage[cp1251]{inputenc}

\usepackage[eqsecnum]{cmpj3}

\usepackage{amsmath}
\usepackage{bbm}
\newcommand{\be}{\begin{equation}}
\newcommand{\ee}{\end{equation}}
\newcommand{\bel}[1]{\begin{equation}\label{#1}}
\newcommand{\bea}{\begin{eqnarray}}
\newcommand{\eea}{\end{eqnarray}}
\newcommand{\balign}{\begin{align}}
\newcommand{\ealign}{\end{align}}
\newcommand{\ba}{\begin{array}}
\newcommand{\ea}{\end{array}}

\def\Z{\mathbb Z}

\def\det{\mathrm{det}}
\def\Tr{\mathrm{\,tr}}
\def\inf{\mathrm{inf}}
\def\id{{\mathbbm{1}}}
\def\det{\textrm{det}}
\def\Hs{\mathfrak H}

\def\i{\textrm{i}}

\issue{2023}{26}{1}{13502}
\doinumber{10.5488/CMP.26.13502}

\title[Local thermalization of  nonequilibrium extended quantum systems]{Some speculations about local thermalization of  nonequilibrium extended quantum systems}

\author[M. Coppola, D. Karevski]{M. Coppola\orcid{0000-0003-0368-5802}, D. Karevski\orcid{0000-0003-4158-3823} \thanks{\email{dragi.karevski@univ-lorraine.fr}.
It is with great pleasure that I wish to contribute to this special edition in honor of my dear friend Bertrand Berche, with whom I have been traveling around the world, visiting Ukraine many times and maintained a friendship for  thirty years now. May this friendship last forever. I also want to dedicate this paper to the memory of our mentor Lo\"ic Turban who supervised our phds and who guided us during all these years.  
} }
      
\address{
 Universit\'e de Lorraine, CNRS, LPCT, F-54000 Nancy, France
}

\date{Received November 14, 2022, in final form December 6, 2022}

\begin{document}

\maketitle

\begin{abstract}
We discuss the possibility of defining an emergent local temperature in extended quantum many-body systems 
evolving out of equilibrium. For the most simple case of free-fermionic systems, we give an explicit formula  for the effective temperature in the case of, not necessarily unitary, Gaussian preserving dynamics. In this framework, we consider the hopping fermions on a one-dimensional lattice submitted to randomly distributed projective measurements of the local occupation numbers. We show from the average over many quantum trajectories that the effective temperature relaxes exponentially towards  infinity. 
%
%
%(\href{https://physh.aps.org/browse}{Physics Subject Headings})
\printkeywords
%
%\pacs Up to six PACS numbers (optional)
\end{abstract}

\section{Introduction}
In the last decade, there has been a tremendous  interest in the study of the out of equilibrium properties of simple many-body systems in both unitary and non-unitary cases. 
In the unitary settings, the out of equilibrium dynamics is generated by the more or less sudden quench of some global \cite{Calabrese2005,Amico}
or local \cite{Eisler2007,Eisler2008} Hamiltonian parameter, after which the system is left to evolve unitarily in time.
The quench may be also homogeneous or inhomogeneous, varying the local couplings at different rates \cite{Collura2010,Collura2011,Scopa2017,Scopa2018}. 
 In this scenario, one important way of catching the non-equilibrium properties of the system is by considering the dynamical evolution of the entanglement spreading all over the system \cite{Eisert2010,Laflorencie}. For local short-range Hamiltonians, starting from a disentangled initial state, it has been shown that the spreading of the quantum correlations is a ballistic process governed by the existence of a Lieb-Robinson bound \cite{LiebRobinson}. From each points of the system, after the quench, correlation fronts propagate at a maximum velocity, the Lieb-Robinson bound, and give rise to local  light-cones effects. 
For integrable one-dimensional systems, with infinite-life time quasi-particle excitations 
propagating ballistically, the entanglement spreading is linear in time, as it can be easily understood from a simple 
quasi-particle picture \cite{Calabrese2005,Alba2018}. 
At later times, the system relaxes locally to a generalized Gibbs state which, in general, is defined by the knowledge of an infinite set of conserved quantities, reflecting an extensive scaling of the entanglement entropy \cite{Rigol2008,Ilievski2015,Ilievski2016,Abanin2019}. 
However, many factors may alter the unitarity of the evolution: in real life, the system is never perfectly isolated and it can also be monitored non-unitarily as it is the case when projective measurements are performed \cite{Wiseman}.
Under such a monitoring, the dynamics may be drastically modified, leading for example to the possibility of suppressing completely the spreading of the entanglement \cite{Cao,Skinner,Alberton,Coppola}. 
The non-unitary dynamics in many situations can be described by a Markovian dynamical map, leading to a Lindblad time evolution equation for the density matrix of the system \cite{Breuer,Karevskibook}, instead of the unitary Liouville equation. The opening of the quantum system may also be global, where dissipative channels are connected to the whole system, or may be  local, where  some regions of the system only are governed  by a dissipative map while the remaining  part is governed by a Hamiltonian dynamics. 
One such a local situation is when a finite chain is coupled at its ends to different baths (or reservoirs of particles), generating at large enough times a non-equilibrium steady state (NESS)  \cite{Karevski2009,Platini2010,Prosen2011a,Prosen2011b,Karevski2013,Popkov,Karevski2014,Landi2015}.  The local quantum NESS generated this way is a stationary
current full quantum state which is given by a generalized MacLennan-Zubarev state \cite{MacLennan,Zubarev}, which can also be called a generalized Gibbs state, $\rho(x) \sim \re^{-\beta H(x) -\delta J(x)}$, where $\beta= ({\beta_L + \beta_R})/{2} $ is the average inverse temperature of the two baths coupled to the local Hamiltonian $H(x)$ and $\delta=\beta_L -\beta_R$ is the drift parameter coupled to a local current operator $J(x)$ \cite{Karevski2009}. The melting of an initial domain wall may also be described in this way since it represents locally the same physical setup \cite{Antal1,Antal2,Antal3,Aschbacher,Ogata1,Ogata2,Bernard2012,Collura1,Platini1,Karevski0,Platini2,Landi1,Scopa0}.

In this paper we reconsider the situation of a one-dimensional quantum system evolving according to a dynamical map which may or may not be unitary and wish to assign to it, in the most simple possible way, a local temperature. This attempt has been considered in the equilibrium case, where the local temperature of a quantum system, if not properly defined,  may not be a well defined quantity in the sense that it may not be anymore an intensive quantity and differs from the whole system temperature~\cite{Hartmann1,Hartmann2,Hartmann3,Eisler,Garcia,Ferraro,Kliesch,DePasquale}. 
Let us mention that several attempts to define the effective temperatures appeared quite recently in the context of 
the system-plus-reservoir paradigm in the strong coupling limit \cite{Moreno2019}, by the introduction of a virtual temperature in the context of quantum machines \cite{Brunner2012}, or based on the analyses of the heat flow in open quantum systems \cite{Morozov2018,Latune2019}.

In the next section we define the model and its dynamics. Section~\ref{sec3} is devoted to the question of local effective temperatures. We end the paper by some discussion in section~\ref{sec4}.

\section{Out of equilibrium dynamics of an extended quantum system}
\subsection{Local state}
\subsubsection{Hydrodynamical description}

We consider one-dimensional free-fermionic systems defined on an infinite one-dimensional lattice with lattice spacing $a$. The system may also be submitted to an external potential  $V(x)$ which is assumed to be a real  smooth function. 
To describe the local properties of the system, we split the one-dimensional lattice into regular intervals $[x, x+ a\ell[$, each containing   a large number $\ell$ of lattice sites, while keeping the width $a\ell$ of the cell small enough such that for whatever sites $j\in[x, x+ a\ell[$ the value of the potential is constant in that interval: $V_j\simeq V(x)$. The Hilbert space of the system is thus decomposed as a product of  identical finite-dimensional  local Hilbert spaces  $\Hs_x $ located at position $x$. The Hamiltonian $H= \int \rd x H(x)$ of the full system is thus a sum over the real line of the Hamiltonian density $H(x)$ which in the hydrodynamic limit, for example for spinless fermionic tight binding models, is given by \cite{Karevski1,Karevski2,Scopa1,Scopa2,Riggio}
\be
H(x) = \frac{1}{a\ell}  \int_{a\ell} \rd y \;  \left\{ \kappa(x) \left( \psi_x^\dagger(y) \psi_x(y+a) + \psi_x^\dagger(y+a) \psi_x(y)\right) + V(x) \psi_x^\dagger(y)\psi_x(y) \right\},
\ee
where $\kappa(x)$ is the local hopping constant and the $\psi_x$ and $\psi_x^\dagger$ are Fermi fields defined in the hydrodynamical  cell $x$. 
Within each coarse-grained point $x$, the system can be diagonalized by a Fourier-Bogoliubov transformation.
Obviously, in order to make this continuum limit practicable, the local scale $\ell$ should be much larger than the inverse particle density: $\ell \gg 1/\rho$. In any case, one can always go back to the discret nature of the system and use the discret splitting $H= \sum_x H(x)$, where $H(x)$ is the local Hamiltonian defined on the lattice sites $j\in [x, x+ a\ell[$
\begin{equation}
H(x)=c^\dagger(x) T(x,0) c(x) = \sum_{i,j\in x}  c^\dagger_i(x) T_{ij}(x,0) c_j(x) . 
\label{localH}
\end{equation} 
The $c^\dagger_i(x), c_{i}(x)$ are local creation and annihilation operators satisfying the usual canonical rules 
\begin{equation}
\{ c^\dagger_i(x),c_j(x')\} = \delta_{ij} \delta_{xx'}, 
\end{equation}
and $T(x,0)$ is the local one-particle Hamiltonian ($\ell\times \ell$ Hermitian coupling matrix). 

If the hydrodynamical description is valid, the lattice variables $c^\dagger_i(x), c_{i}(x)$ may be replaced by the continuous creation and annihilation fields $\psi^\dagger_x(y)$ and $\psi_x(y)$ acting at position $y$ within the coarse-grained cell $x$, such that
\begin{equation}
H(x) = \psi^\dagger_x T(x,0) \psi_x= \int_{\Delta x} \rd y  \int_{\Delta x}  \rd y' \;  \psi^\dagger_x(y) T_x(y,y') \psi_x(y') , 
\end{equation}
where $T_x(y,y')$ is the kernel function associated to the $T(x,0)$.

\subsubsection{Time evolution of the local state}
The local initial state at position $x$ is deduced from the initial global state $\rho(0)$ by tracing out the degrees of freedom living in the complementary space $\Hs_{/x}$ of the local Hilbert space  $\Hs_x$:
\begin{equation}
\rho(x,0) = \Tr_{ \Hs_{/x}} \{ \rho(0)\}\, . 
\end{equation}
Up to boundary corrections, we assume that this local state is initially given by a local canonical state\footnote{For a general argument why the canonical states are ubiquitous states see \cite{Popescu,Goldstein}.}
\begin{equation}
\rho(x,0)=\omega\left( H(x),\beta(x)\right) \equiv\frac{\re^{-\beta(x) H(x)}}{{Z}[H(x),\beta(x)]} \,,
\label{rhogibbs}
\end{equation}
where   $H(x)$ is the local Hamiltonian (\ref{localH}) and ${Z}(H,\lambda)=\Tr_{\Hs_x}\{ \re^{-\lambda H(x)})\}$ is the local  partition function.

At a later time $t$, the system  evolves according to a dynamical map $\Lambda_t$ such that 
\begin{equation}
\rho(t) = \Lambda_t [ \rho(0)],
\end{equation}
from which the local state is obtained by taking the partial trace 
\begin{equation}
\rho(x,t) =  \Tr_{ \Hs_{/x}} \{ \rho(t)  \} = \Tr_{ \Hs_{/x}} \{ \Lambda_t [\rho(0)]\} \; . 
\end{equation}
If the global dynamical map $\Lambda_t$ is unitary, then 
\begin{equation}
\rho(t) = \Lambda_t [ \rho(0)] = U(t,0) \rho(0) U^\dagger(t,0),
\end{equation}
where $U(t,0)$ is the unitary time evolution operator defined on the global Hilbert space $\Hs$. 
In such a case, the local state is given by
\begin{equation}
\rho(x,t) =  \Tr_{ \Hs_{/x}} \left\lbrace  U(t,0) \rho(0) U^\dagger(t,0) \right\rbrace  . 
\end{equation}
In general, the system is not perfectly isolated and/or subject to an external monitoring, such as discret or continuous  driving of some parameter, measurements of some local properties, and this in general breaks the unitarity of the dynamical map $\Lambda_t$ \cite{Coppola}. 

However, even for a perfectly isolated system, with a global unitary dynamics, the time evolution of the reduced density matrix, that is the local state $\rho(x,t)$, is not in general a unitary evolution \cite{Breuer,Karevskibook}. 
Indeed, if we assume for simplicity that the initial global state $\rho(0)$ is a tensor product state of the $x$-cell state times, the state $\omega_e$ of the remaining part, that is $\rho(0)= \rho(x,0) \otimes \omega_e(0)$, then the local state at a later time is given by 
\begin{equation}
\rho(x,t) = \Tr_{ \Hs_{/x}} \left\lbrace  U(t,0) \rho(x,0)\otimes \omega_e(0) U^\dagger(t,0) \right\rbrace  . 
\end{equation}
Now, introduce the spectral decomposition of the environement state  $\omega_e(0) = \sum_k \lambda_k \pi_k$, where $\{\pi_k =  |f_k\rangle \langle f_k| \}$ is a complete family of orthogonal projectors in the Hilbert space $\Hs_{/x}$, and  the tensor decomposition of the unitary evolution operator
\begin{equation}
U(t,0) = \sum_{ij} W^{ij}(t) \otimes M^{ij},
\end{equation}
where $M^{ij} = |f_i\rangle \langle f_j |$ is defined on the environement and 
$W^{ij}(t)$ is a matrix defined on $\Hs_x$ which can be decomposed on  
an orthonormal basis $\{|\varphi_k\rangle\}$ of the local space $\Hs_x$ as
$W^{ij}(t) = \sum_{kl} (W^{ij}(t))_{kl} L^{kl}$ with $L^{kl} = |\varphi_k\rangle \langle \varphi_l |$.
Plugging that into the time-evolved reduced density matrix one obtains
\begin{equation}
\rho(x,t) = \sum_k \lambda_k \sum_i W^{ik}(t) \rho(x,0) {W^{ik}}^\dagger(t) \; .
\end{equation}
Since the $\lambda_k$ are the eigenvalues of the environment density operator, they are all positive and one can redefine the Kraus operators as $K^{ij} = \sqrt{\lambda_j} W^{ij}$ such that the reduced dynamical map is given by
\begin{equation}
\rho(x,t) = \Lambda_{(x,t)}[\rho(x,0)] = \sum_\alpha K^\alpha(t) \rho(x,0) {K^\alpha}^\dagger(t),
\end{equation}
with the trace preserving condition 
\begin{equation}
\sum_\alpha {K^\alpha}^\dagger(t)  K^\alpha(t)  = \id_{\Hs_x} . 
\end{equation}
When there is only one Kraus operator left, the local dynamics is unitary too. 

If the local dynamical map $\Lambda_{(x, t)}$ satisfies the Markovian semi-group property
\begin{equation}
\Lambda_{(x,t)} \Lambda_{(x,s)} = \Lambda_{(x,t+s)},
\end{equation}
then 
\begin{equation}
\rho(x,t) = \Lambda_{(x,t)} \rho(x,0) = \Lambda_{(x,\epsilon)}\Lambda_{(x,t-\epsilon)} \rho(x,0) =  \Lambda_{(x,\epsilon)} \rho(x,t-\epsilon)   ,
\end{equation}
from which one deduces that there is an infinitesimal generator ${\cal L}_x$, such that the infinitesimal dynamical map $\Lambda_{(x,\epsilon)} = \id + \epsilon {\cal L}_x$, whose exponential gives the finite dynamics:
\begin{equation}
\Lambda_{(x,t)} = \lim_{n\rightarrow \infty} \left(\id + \frac{t}{n} {\cal L}_x\right)^n =\re^{t {{\cal L}_x}}  . 
\end{equation}
Taking the time derivative of $\rho(x,t) = \re^{t {{\cal L}_x}} \rho(x,0)$ one has the differential equation
\begin{equation}
\partial_t \rho(x,t) = {\cal L}_x \rho(x,t) ,  
\end{equation}
which can be written in the Lindblad form 
\begin{equation}
\partial_t \rho(x,t) = -\ri[H(x),\rho(x,t)] + \sum_{k} L_k(x) \rho(x,t) L_k^\dagger(x) -\frac{1}{2} \left\{L_k^\dagger(x) L_k(x), \rho(x,t)\right\},
\end{equation}
where $H(x)$ and  the so-called jump operators $\{L_k(x)\}$  acting on the local Hilbert space $\Hs_x$ are defined  in terms of the infinitesimal Kraus operators $\{K^\alpha(\epsilon)\}$ under a proper scaling limit \cite{Breuer,Karevskibook} and with $\{A,B\} \equiv AB + BA$ the anticommutator of $A$ and $B$. 

Solving the Lindblad equation for a many-body system is in general a very difficult task, even if a few exact results do exist for non-interacting particles \cite{Karevski2009,Prosen2008} and for some Bethe-integrable systems \cite{Prosen2011b,Karevski2013,Popkov,Prosen2015}.

\subsection{Gaussian preserving dynamics\label{sectionGPD}}
In this context, we consider a special class of dynamical maps that preserves the Gaussianity of the local state. 
Gaussian states $\rho(x)$ are states for which Wick theorem applies and they are
fully characterized by the correlation matrix (two-point functions)
\be
C_{ij}(x) = \langle c^\dagger_i c_j \rangle_{\rho(x)} = \Tr_{\Hs_x} \left\{ c^\dagger_i c_j \rho(x) \right\}\; . 
\label{Cij}
\ee
If the initial state $\rho(x,0)$ is Gaussian, then the unitarily evolved state $\rho(x,t) = \re^{\i tH(x)} \rho(x,0) \re^{-\i tH(x)} $ remains Gaussian and is fully characterized by $C_{ij}(x,t)  = \Tr_{\Hs_x} \{ c^\dagger_i c_j \rho(x,t) \}$. 

However,  Gaussian Preserving Dynamics (GPD) are more general and can be implemented by other schemes. 
One of them is by a continuous monitoring of the system under projective measurements of some local densities \cite{Coppola}, which obviously destroys, in general, the unitarity of the dynamics. To be more specific, consider a local observable $Q_\Omega$, defined on some compact support $\Omega \subset \Z$, 
\be
Q_{\Omega} = \sum_p q_p P^{(p)}_\Omega  , \quad \sum_p P^{(p)}_\Omega = \id_\Omega  ,
\ee
where the $P^{(p)}_j$ are the orthogonal projectors on the corresponding subspace associated to the eigenvalue $q_p$ of the observable $Q_\Omega$.
Starting, for example, with a pure  state $|\Psi\rangle$, 
just after the measurement of the local observable $Q_\Omega$ with an outcome $q_k$, the state is projected according to the Born rule
\be
|\Psi \rangle \longrightarrow \frac{ P^{k}_\Omega |\Psi\rangle }{\langle \Psi | P^{k}_\Omega |\Psi\rangle} \; . 
\ee
In general, projective measurements will not preserve the Gaussianity of the state. However, 
if one considers the measurements of the local density $\hat{n}_j = c^\dagger_j c_j$,
with the two possible outcomes $q_0=0$ and $q_1=1$,
 then the dynamics remains GPD. 
Indeed, the local density operator can be represented by $\hat{n}_j =  q_1 P_j^{(1)} + q_0 P_j^{(0)}$ with $P_j^{(1)} +  P_j^{(0)} = \id_j$ which implies that all local number operators are projectors: 
\be
\hat{n}_j =  P_j^{(1)}\; , \quad \id_j-\hat{n}_j =   P_j^{(0)}  . 
\ee
Using the operator identity
\be
\re^{\nu \hat{n}_j } = \id_j + (\re^\nu -1) \hat{n}_j \, , 
\ee
we see that the projectors $P_j^{(0)}=\id_j-\hat{n}_j $ and  $P_j^{(1)}=\hat{n}_j  $ can be expressed as limits of Gaussian operators:
\be
\id_j - \hat{n}_j = \lim_{\nu \rightarrow \infty} \re^{- \nu \hat{n}_j }  , \quad 
\hat{n}_j =  \lim_{\nu \rightarrow \infty} \frac{\re^{\nu \hat{n}_j }}{\re^\nu -1}
 . 
\ee
Since by hypothesis the initial state is Gaussian, $\rho \propto \re^{\sum_{ij}c^\dagger_i M_{ij} c_j} $, from the projection rules associated to either $\id_j - \hat{n}_j $ or $\hat{n}_j $  one has  to consider the expression
\be
\re^{\pm \nu \hat{n}_j } \re^{\sum_{kl}c^\dagger_k M_{kl} c_l} \re^{\pm \nu \hat{n}_j },
\ee 
which, from Baker-Campbell-Hausdorff formula, is nothing else but a Gaussian state
\be
e^{\sum_{kl}c^\dagger_k K_{kl} c_l} ,
\ee
associated to some new coupling matrix $K$. Notice that since a Gaussian state is fully characterized by its correlation matrix $C_{ij}$, the projection rules can be translated into the following rules for the two-point functions:
If the outcome of the measurement of the local density at site $k$ is $1$, which occurs with probability $p_k=C_{kk}(x,t)=\langle \hat{n}_k \rangle$, then
\begin{equation}
C_{ij}(x,t) \longrightarrow \delta_{ik}\delta_{jk} + C_{ij}(x,t) - \frac{C_{ik}(x,t) C_{kj}(x,t)}{C_{kk}(x,t)} 
\label{Cij1}
\end{equation}
and otherwise
\begin{equation}
C_{ij}(x,t) \longrightarrow -\delta_{ik}\delta_{jk} + C_{ij}(x,t) + \frac{[\delta_{ik} - C_{ik}(x,t)] [\delta_{jk}- C_{kj}(x,t)]}{1-C_{kk}(x,t)} \; . 
\label{Cij0}
\end{equation}
Indeed, when the outcome of the measurement of the density at site $k$ is 1, the projection rule transforms the two-point function $C_{ij} = \Tr\{c_i^\dagger c_j \rho\}$ into 
$\Tr\{c_i^\dagger c_j \hat{n}_k  \rho \hat{n}_k  \}= \Tr\{ \hat{n}_k c_i^\dagger c_j \hat{n}_k  \rho  \} = \langle c^\dagger_k c_k c_i^\dagger c_j   c^\dagger_k c_k \rangle$. 
After normal ordering this six-point function, one obtains $\delta_{ik}\delta_{jk} \langle c^\dagger_k c_k\rangle + 
 \langle c^\dagger_k c^\dagger_i c_j c_k\rangle$ and the last term is reduced to 
 $\langle c^\dagger_k c_k\rangle \langle c^\dagger_i c_j\rangle  -\langle c^\dagger_k c_j\rangle \langle c^\dagger_i c_k\rangle $
 thanks to Wick theorem. After dividing $\Tr\{c_i^\dagger c_j \hat{n}_k  \rho \hat{n}_k  \}$ by the proper normalization factor $\Tr\{ \hat{n}_k \rho \hat{n}_k\} = \Tr\{ \hat{n}_k \rho\} = C_{kk}$, one obtains the projection rule \eqref{Cij1}. Following the same logic one obtains the rule \eqref{Cij0}.

Whenever no measurement
occurs, the time evolution is unitary and GPD such that the correlation matrix $C$ evolves according to $C(t+\tau)= R^\dagger(\tau) C(t) R(\tau)$, where $R(\tau)$ is a unitary matrix. Consequently, the combination of the unitary evolution and these projection rules leads to a quantum trajectory of the system which is governed by a non-unitary GPD \cite{Coppola}.

\section{Local effective temperature}\label{sec3}
\subsection{Evolution of the coupling matrix}
Consider now a system initially prepared in a state such that the local density operator is a Gibbs state~(\ref{rhogibbs}) at some local inverse temperature $\beta(x)$.
We assume that at a later time the new local state generated by the local (non-unitary) dynamical map  $\Lambda_{(x,t)}$ remains gaussian: 
\begin{equation}
\rho(x,t)=\frac{1}{Z[H(x,t),\beta(x)]} \re^{-\beta(x) c^\dagger(x) T(x,t) c(x)},
\label{gauss1}
\end{equation}
where the normalization factor $Z[H(x,t),\beta(x)] = \Tr_{\Hs_x} \left\lbrace  \re^{-\beta(x) c^\dagger(x) T(x,t) c(x)}\right\rbrace $ and  with the new one particle matrix 
\be
T(x,t) = \lambda_{(x,t)} [T(x,0)] 
\ee
evolved non-unitarily from the initial one particle Hamiltonian matrix $T(x,0)$ with a dynamical map $ \lambda_{(x,t)}$. 
One can relate the dynamical evolution of the coupling matrix $T(x,t)$ to the dynamical evolution of the correlation matrix $C(x,t)$ due to the relation 
\be
C^{\text{tr}}(x,t) = \frac{ 1}{\id_x+\re^{\beta(x)  T(x,t)}} \quad  \Leftrightarrow \quad  -\beta(x) T(x,t) = \ln \frac{C^{\text{tr}}(x,t)  }{\id_x-C^{\text{tr}}(x,t) },
\ee
where the upper script $\text{tr}$ stands for the transposed  matrix and $\id_x$ is the $\ell\times \ell$ identity matrix defined on the $x$ cell.

Even if the quadratic form 
\begin{equation}
H(x,t)\equiv c^\dagger(x) T(x,t) c(x)
\label{Ht}
\end{equation}
entering  the local state $\rho(x,t)$ 
 may be interpreted as a new Hamiltonian, one is not legitimate to interpret $\rho(x,t)$ as a Gibbs state since 
 $H(x,t)$ is not the true local system Hamiltonian $H(x)$. 
 
 Let $W_t(x)$ be the unitary matrix  diagonalizing the coupling matrix $T(x,t)$:
 \begin{equation}
 W_t^\dagger(x) T(x,t) W_t(x) = E(x,t), 
 \end{equation}
 where $[E(x,t)]_{qp} = \epsilon_p(t) \delta_{qp}$ is the diagonal matrix associated to $T(x,t)$. The quadratic form $H(x,t)$ may then be expressed in terms of instantaneous diagonal Fermi operators $\eta(x,t) = W_t^\dagger(x) c(x)$ such that 
\begin{align}
H(x,t) &= c^\dagger(x) T(x,t) c(x) = c^\dagger(x) W_t(x)  E(x,t) W_t^\dagger(x)c(x) \nonumber \\
&= \eta^\dagger(x,t) E(x,t) \eta(x,t) = \sum_q \epsilon_q(t) \eta^\dagger_q(x,t) \eta_q(x,t) \; . 
\end{align}
 The relation between the instantaneous diagonal Fermi operators and the diagonal Fermi operators (at $t=0$) of the local Hamiltonian $H(x) = \sum_q \epsilon_q(x) \eta_q^\dagger(x) \eta_q(x)$ is obtained from
 \begin{equation}
 \eta(x,t) = W_t^\dagger(x) c(x) =  W_t^\dagger(x) W_0(x) \eta(x) \equiv D_t(x) \eta(x)\; , 
 \end{equation}
 where $D_t(x) = W_t^\dagger(x) W_0(x) $ defines a unitary mapping transforming  $\eta(x)$ into $\eta(x,t)$. The instantaneous Hamiltonian (\ref{Ht}) can thus be written as 
\begin{align}
H(x,t)\equiv c^\dagger(x) T(x,t) c(x) = \eta^\dagger(x) D_t^\dagger(x) E(x,t) D_t(x) \eta(x) \equiv  \eta^\dagger(x) \Omega(x,t)  \eta(x)  \; , 
\end{align} 
where the new matrix 
\begin{equation}
\Omega(x,t) = D_t^\dagger(x) E(x,t) D_t(x) = W_0^\dagger(x) T(x,t) W_0(x)  
\label{Omegamatrix}
\end{equation}
is not in general  diagonal. With the matrix elements 
$[\Omega(x,t) ]_{qp} = \omega_{qp}(x,t)$, the explicit expression of $H(x,t)$ is: 
\begin{align}
H(x,t) = H(x) +
\sum_q \left[ \omega_{qq}(x,t) -\epsilon_q(x)\right] \eta_q^\dagger(x) \eta_q(x) + \sum_{q\neq p}  \omega_{qp}(x,t) \eta_q^\dagger(x) \eta_p(x) \; . 
\end{align}
The diagonal correction to $H(x)$ describes the energy level shift of the original single-particle spectrum while the non-diagonal term makes the transitions between the single-particle states.

\subsection{Effective local temperature}
Assuming that the system is at time $t$ in a local Gaussian state of the form (\ref{gauss1}), we want to quantify how close is that state to an actual local canonical state $\omega(H(x),\lambda)$ at inverse temperature $\lambda$. 
 If the local state is very close to this canonical state,  we shall then identify the parameter $\lambda$ as the effective local inverse temperature  $\beta(x,t)$ of the system at position $x$ and time $t$.

To properly quantify the closeness of the states, we consider the distance 
\begin{equation}
{d}(x,t,\lambda)\equiv \|\rho(x,t)-\omega(H(x),\lambda)\| \; 
\label{distance}
\end{equation}
defined from the Hilbert-Schmidt operator norm (on bounded operators) 
\begin{equation}
\|A\|=\sqrt{\Tr\{A^\dagger A\}} \; .
\end{equation}
A more natural measure of the distance between two states $\rho_1$ and $\rho_2$ would have been given by the trace-norm $\|\rho_1-\rho_2\|_1$, where
$$
\|A\|_1=\Tr| A| = \Tr\sqrt{A^\dagger A}\; , 
$$
as this distance is small if and only if the two states are effectively indistinguishable in the experimental sense. This is not the case for the Hilbert-Schmidt norm since, for high-dimensional spaces,  it can be small even for perfectly distinguishable states (for example orthogonal states). Nevertheless,   the two norms are related through the inequality
$$
\|A\|_1\leqslant \sqrt{d} \|A\|,
$$ 
where $d$ is the dimension of the Hilbert space on which $A$ is defined. 
In this sense, for finite-dimensional spaces, the two norms lead to physically equivalent conclusions. The advantage of using the Hilbert-Schmidt norm is that it is much easier to be handled.

The effective local inverse temperature  $\beta(x,t)$ at time $t$ is defined through the minimization of the distance $d(x,t,\lambda)$ over the one-parameter canonical family  $\omega(H(x),\lambda)$ :
\begin{equation}
d(x,t,\beta(x,t))\equiv \inf_{\lambda} d(x,t,\lambda)\; .
\end{equation}
Taking  the $\lambda$ derivative of $d^2(x,t,\lambda)$ one arrives at 
\begin{equation}
\partial_\lambda d^2(x,t,\lambda) = 2 \langle \left(  H(x) - \langle H(x) \rangle_{\omega(H(x),\lambda)}\right)\left(\rho(x,t) -\omega(H(x),\lambda) \right) \rangle_{\omega(H(x),\lambda)},
\end{equation}
where we have defined the  expectation value $\langle A \rangle_{\omega} \equiv \Tr\{ A \omega  \}$. 
Consequently, the roots of that equation are simply given by the vanishing of the connected correlation function 
\begin{equation}
\langle H(x)\Delta_\lambda \rho(x,t)\rangle_\lambda^c \Big|_{\lambda=\beta(x,t)}=0 ,
\label{mini}
\end{equation}
where $\Delta_\lambda\rho(x,t)\equiv \rho(x,t)-\omega(H(x),\lambda)$ and
$\langle AB\rangle^c = \langle AB\rangle -\langle A\rangle\langle B\rangle $. See \cite{DePasquale,Kliesch} for a somehow  related approach based on the fidelity measure. 

To give an explicit formula in the case of Gaussian states, we need to compute the following trace
\begin{equation}
\Tr\left\lbrace  \re^{c^\dagger(x) A(x) c(x) } e^{c^\dagger(x) B(x) c(x) } \right\rbrace  . 
\end{equation}
Noticing that 
\begin{equation}
[c^\dagger(x) A(x) c(x) , c^\dagger(x) B(x) c(x) ] = c^\dagger(x) [A(x),B(x)] c(x)
\end{equation}
one has from Backer-Campbell-Hausdorff formula
\begin{equation}
\re^{c^\dagger(x) A(x) c(x) } \re^{c^\dagger(x) B(x) c(x) }  = \re^{c^\dagger(x) A(x) c(x) + c^\dagger(x) B(x) c(x) + \frac{1}{2} [c^\dagger(x) A(x) c(x) , c^\dagger(x) B(x) c(x) ] + \dots}\,,
\end{equation}
that is, using the previous commutator identity,
 \begin{equation}
\re^{c^\dagger(x) A(x) c(x) } \re^{c^\dagger(x) B(x) c(x) }  = \re^{c^\dagger(x) \left( A(x) + B(x) + \frac{1}{2}[A(x),B(x)]+\dots\right) c(x)} = \re^{c^\dagger(x) M(x) c(x)}\,,
\end{equation}
with the matrix $M$ defined by
\begin{equation}
\re^{M(x)} = \re^{A(x)} \re^{B(x)}\; . 
\end{equation}
Now, since the trace of a Gaussian state is 
\begin{equation}
\Tr\left\lbrace \re^{c^\dagger(x) M(x) c(x)} \right\rbrace   = \textrm{det}(\id_x+ \re^{M(x)}) \; , 
\end{equation}
where $\id_x$ is the unit matrix of size $\ell$ (the size of the local cell), one arrives at the identity
\begin{equation}
\Tr\left\lbrace \re^{c^\dagger(x) A(x) c(x)} \re^{c^\dagger(x) B(x) c(x)}  \right\rbrace   = \textrm{det}(\id_x+  \re^{A(x)} \re^{B(x)} ) \; . 
\end{equation}
Together with the formula
\begin{equation}
\partial_\lambda \det (\id+ \re^{\lambda A} \re^B) = \det(\id+ \re^{\lambda A } \re^B) \Tr\{ (\id+ \re^{\lambda A} \re^B)^{-1} A \re^{\lambda A} \re^B \}
\end{equation}
and (\ref{gauss1}), the equation (\ref{mini}) leads to 
\begin{align}
&\frac{ \det(\id_x + \re^{-2\lambda T(x,0)}) } { \det(\id_x + \re^{-\lambda T(x,0)} )} 
\,  \Tr \left\{ T(x,0) 
 \left( \frac{1}{\id_x + \re^{\lambda T(x,0)} }- \frac{1}{\id_x + \re^{2\lambda T(x,0)} } \right) 
 \right\}
\nonumber \\
&- \left.  \frac{ \det(\id_x + \re^{-\lambda T(x,0)} \re^{-\beta(x) T(x,t)} ) } { \det(\id_x + \re^{-\beta(x) T(x,t)} )} 
\right.\nonumber \\  
& \left.\times\Tr \left\{ T(x,0) 
 \left( \frac{1}{\id_x + \re^{\lambda T(x,0)} }- \frac{1}{\id_x + \re^{\beta(x) T(x,t)} \re^{\lambda T(x,0)} } \right) 
 \right\} \right|_{\lambda=\beta(x,t) } =0   \; . 
\end{align}
This cumbersome explicit condition is the main result we wanted to derive. 

At high temperature initial state and assuming that the effective local temperature remains high too, that is for $\lambda \sim \beta(x) \ll 1$, using the identity $\det( \id_x + \epsilon M) = 1+\epsilon  \Tr\{ M\}  + o(\epsilon)$,  to the leading order in $\beta(x)$, the previous equation drastically simplifies into
\begin{equation}
\left. \lambda \Tr\left\{ T^2(x,0)\right\} - \beta(x) \Tr\left\{ T(x,0) T(x,t)\right\}  \right|_{\lambda = \beta(x,t)} =0 , 
\end{equation}
which gives the effective local temperature as
\begin{equation}
\beta(x,t) = \beta(x) \frac{ \Tr\{ T(x,0) T(x,t)\} } { \Tr\{ T^2(x,0)\} }    =  \beta(x)  \left( 1+ \frac{ \Tr\{T(x,0)\Delta(x,t) \}  }{\Tr\{ T^2(x,0) \}} \right),
\label{beta1}
\end{equation}
where we have defined the difference matrix $\Delta(x,t) \equiv T(x,t)-T(x,0)$. Using (\ref{Omegamatrix}) we have in terms of the one-particle spectrum
\begin{equation}
\beta(x,t) = \beta(x) \frac{ \Tr\{ E(x,0) \Omega(x,t)\} } { \Tr\{ E^2(x,0)\} }  = \beta(x) \frac{ \sum_q \epsilon_q(x) \omega_{qq}(x,t) } {\sum_q \epsilon_q^2(x) }\; .
\end{equation}

In order to appreciate the meaning of this formula, let us introduce the scalar product defined on  $\textrm{End}(\Hs_x)$ by 
\begin{equation}
(A,B) = \Tr\{A^\dagger B\}\; , \quad \forall  \; A,B \in \textrm{End}(\Hs_x)\; . 
\end{equation}
Equation (\ref{beta1}) can thus be written as 
\begin{equation}
\beta(x,t) = \beta(x) \frac{ (T(x,0), T(x,t)) }{(T(x,0), T(x,0))} \; . 
\label{beta2}
\end{equation}
This means that only the non-orthogonal part of the actual one-particle Hamiltonian $T(x,t)$, with respect to the chosen reference state defined through the one-particle Hamiltonian $T(x,0)$, contributes to the effective local inverse temperature $\beta(x,t)$. Whatever part in $T(x,t)$ that is orthogonal to $T(x,0)$ will not affect the value of the local inverse temperature, it may in general nevertheless increase the distance to the optimum canonical state $\omega(H(x),\beta(x,t))$.
If the  one-particle Hamiltonian $T(x,t)$ is completely orthogonal to the coupling matrix of the system, then this procedure will fix an infinite value to the local temperature.  
 
Moreover, since in a hydrodynamical cell of size $\ell$ the local system is translation invariant, the local  Hamiltonian can be 
decomposed into a sum over local conserved charges $H(x)=\sum_k H^{(k)}(x) =  \sum_k q^{(k)}(x) c^\dagger(x)    Q^{(k)}(x) c(x)$, where the matrices $Q^{(k)}(x)$ are given by\footnote{We suppose here that the hydrodynamical cell is large  enough such that we can neglect the boundary details. The $H^{(k)}(x)$ are charges in the sense that they commute with each other and as a consequence with the Hamiltonian $[H(x),H^{(k)}(x)]=0$ which implies that they are conserved quantities $\partial_t H^{(k)}(x)=0$. 
Locally, this translates into  continuity equations satisfied by the charge densities $q^{(k)}(x)  [ c^\dagger_j(x) c_{j+k}(x) + c^\dagger_{j+k}(x) c_{j}(x) ] $.}
\begin{equation}
\left(Q^{(k)}(x) \right)_{ij} =  \frac{1}{2} (\delta_{i,j+k} + \delta_{i+k,j} )
 \end{equation}
and $q^{(k)}(x)$s are real coefficients. 
Typically,  $Q^{(0)}$ gives the on-site energy contribution or potential energy
\begin{equation}
H^{(0)}(x) = q^{(0)}(x) \sum_j c^\dagger_j(x) c_j(x)\; , 
\end{equation}
while $Q^{(1)} (x)$ gives the kinetic contribution (one-site hopping terms) 
\begin{equation}
H^{(1)}(x) = q^{(1)}(x)  \sum_j \left(c^\dagger_j(x) c_{j+1}(x) + c^\dagger_{j+1}(x) c_{j}(x) \right)
\end{equation}
 and so on. In general, usual short range Hamiltonians contain only very few of these charges but one may also encounter long-range hopping situations, for which in general the physical coefficients $q^{(k)}$ are given by a decaying function of the hopping distance $k$, such as a power law $q^{(k)} \sim k^{-\alpha}$ with some positive exponent $\alpha$.  
Notice here that we do not consider in the reference Hamiltonian $H(x)$  the presence of current like terms, associated to antisymmetric Hermitian matrices of the form $\i j^{(k)}(x) (\delta_{i,j+k} - \delta_{i+k,j} ) $  leading to terms like 
$J^{(1)}(x) = \i j^{(1)}(x)  \sum_j \left(c^\dagger_j(x) c_{j+1}(x) - c^\dagger_{j+1}(x) c_{j}(x) \right)$, but in principle we could. 

To be more specific, let us suppose that the reference Hamiltonian is the sum over the $K+1$ first charges $Q^{(k)}$. 
The set of charges $\{Q^{(k)}\}$ for $k=0, 1,\dots , K$ is obviously not a complete set but 
all the charges are orthogonal to each other since $(Q^{k},Q^{k'})\propto \delta_{k,k'}$. 
Indeed, the $k$ matrix $Q^{(k)} = \frac{1}{2} \left( L_k + R_k \right)$, where $L_k$ is the shift to the left by a distance $k$ and $R_k $ the corresponding shift to the right. 
For $k=0$, $Q^{(0)} = L_0=R_0= \id_x$. 
Since $L_k^\dagger=R_k$, one has $(L_k, R_p)= \Tr\{ L_k^\dagger R_p\} = \Tr\{ R_{p+k}\} = 0$ and consequently
\begin{align}
(Q^{(k)}, Q^{(p)}) &= \frac{1}{4} \left[ (L_k,L_p) + (R_k, R_p)\right] = \frac{1}{4} \left( \Tr\{L_{p-k}\} + \Tr\{R_{p-k} \}  \right) =
 \frac{\ell}{2} \delta_{k,p} \quad \forall k\neq 0\; , \\
 (Q^{(0)}, Q^{(0)})  &= \ell\; . 
\end{align}

%$$
%(T(x,0), T(x,0)) = \sum_{k , k'} q^{(k)}(x)  q^{(k')}(x) ( Q^{(k)}(x), Q^{(k')} (x) ) =\left[ \left( q^{(0)}(x)\right)^2 + \frac{1}{2} \sum_{k=1}^K  \left( q^{(k)}(x)\right)^2 \right] \ell
% $$

One can develop the $T(x,t)$ coupling matrix into this set of charges plus a remaining part which is orthogonal to it:
\begin{equation}
T(x,t) = \sum_{k=0}^K q^{(k)}(x,t) Q^{(k)}(x) + T_{\perp} (x,t)\; ,
\label{TTperp}
\end{equation}
with
\begin{align}
(Q^{(p)} (x) , T(x,t)) &=  q^{(p)}(x,t)  \frac{\ell}{2} \quad  \forall p\neq 0 , \\
(Q^{(0)} (x) , T(x,t)) &= q^{(0)}(x,t) \ell \; . 
\end{align}
Notice here that since $Q^{(k)} = \frac{1}{2} \left( L_k + R_k \right)$, the coefficient $(Q^{(k)} (x) , T(x,t)) $ is nothing else but the trace over the two upper and lower  $k$-diagonals of $T(x,t)$, that is $\frac{1}{2} \sum_j [ T_{j,j+k} (x,t) + T_{j+k,j}(x,t) ]$. 
This implies that even if within the cell $x$ the matrix $T(x,t)$ itself is not homogeneous, the outcome to the local thermal properties is   averaged over the hydrodynamical cell. 
Plugging (\ref{TTperp}) into (\ref{beta2}) gives our final formula
\begin{equation}
\beta(x,t) = \beta(x) \frac{  q^{(0)}(x)q^{(0)}(x,t) + \frac{1}{2} \sum_{k=1}^K    q^{(k)}(x) q^{(k)}(x,t)  }
 { \left[ q^{(0)}(x)\right]^2 + \frac{1}{2} \sum_{k=1}^K  \left[ q^{(k)}(x)\right]^2} \; . 
 \label{betaxt}
\end{equation} 
If the Hamiltonian $H(x)$ is a purely hopping Hamiltonian, then only $q^{(1)}(x)$ is non-zero and the formula for the inverse temperature simplifies into
$$
\beta(x,t)= \beta(x) \frac{q^{(1)}(x,t)}{q^{(1)}(x)}\; , 
$$
which is kind of intuitive  since it reflects the simplest identification of $\beta(x,t)  H(x)   \simeq  \beta(x) H(x,t)$, up to orthogonal parts $T_{\perp}(x,t)$, leading in the case of a single charge to  
$\beta(x,t) q^{(k)}(x) Q^{(k)}(x) \simeq  \beta(x) q^{(k)} (x,t) Q^{(k)}(x)  $ that is $\beta(x,t) q^{(k)}(x)  = \beta(x) q^{(k)}(x,t) $.

\subsection{Effective temperature of the hopping fermions with projective measurements of the local densities}

Let us apply this derivation to the  case of the one-dimensional hopping fermions subjected to projective measurements of their local densities that we described in section \ref{sectionGPD}. 
The unitary dynamics is perturbed by the projective measurements at a rate $1/\tau$ of the one-site occupation numbers $\hat{n}_j$. As $\tau$ is increasing, the dynamics is getting closer and closer to the unitary evolution. On the contrary, for a high measurement rate (low $\tau$) the system dynamics is very far from a unitary dynamics, with a very low initial spreading and finally a saturation of the entanglement to a constant value \cite{Coppola}. 
 
The dynamics being GPD, it is fully encoded into the time-evolution of the two-point function \eqref{Cij} through the unitary evolution $C(t+s) = R^\dagger(s) C(t) R(s)$ and the projection rules \eqref{Cij1} and \eqref{Cij0}, from which we reconstruct the coupling matrix $T$. We averaged the deduced inverse temperature \eqref{betaxt} 
over different quantum trajectories\footnote{Different quantum trajectories are generated from an initial state by following in time the different possible outcomes and ``state re-preparations'' that occur after each projective measurements of the local densities.}, 
typically we have taken up to a 100 different trajectories for a cell of size $\ell \sim 100$ sites. 

\begin{figure}[!t]
\centerline{\includegraphics[width=0.9\textwidth]{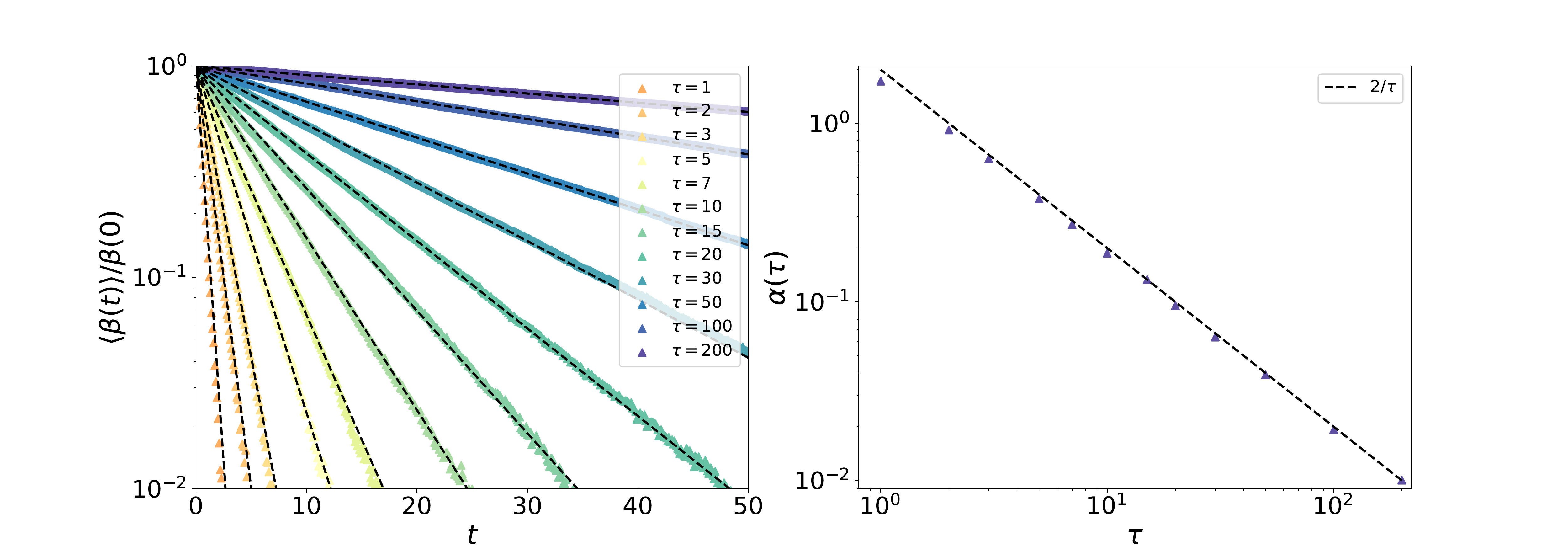}}
\caption{(Colour online) On the left, we show the time evolution of the averaged inverse temperature starting from an initial value $\beta(0)$ (in linear-log scale). The different colored symbols correspond to different values of the measurement rate $1/\tau$. The dashed lines are exponential fits with the ansatz $\langle\beta(t)\rangle/\beta(0)=\re^{-\alpha(\tau)t}$.  
On the right,  we show the evolution of the fitting parameter $\alpha(\tau)$ as a function of $\tau$ in a log-log scale. The dashed line corresponds to $\alpha(\tau) = 2/\tau$. 
}
 \label{Fig1}
\end{figure}

We show the behavior of the effective inverse temperature in figure~\ref{Fig1}. On the left, the scaling of the averaged inverse temperature is plotted versus time for various measurement rates $1/\tau$, computed for several values of the initial inverse temperature from $\beta(0)= 0.1$ up to $\beta(0)= 5$. We observe a clear exponential decay toward zero (infinite temperature) at a rate $\alpha(\tau)$ which is very well fitted by $2/\tau$, see the right-hand figure. This value coincides with the inverse of the average life-time $\tau/2$ of the semi-classical non-interacting quasi-particles that emerge from a Wigner function formalism applied to such a problem \cite{Coppola2}. Within this picture, the average inverse temperature is proportional to the probability that a semi-classical particle, created at the initial time, survives up to time $t$. 
The precise connection of the effective local temperature with the life-time of those semi-classical particles is still under investigation and will be presented in a forthcoming publication.

\section{Discussion}\label{sec4}
We have discussed the possibility of identifying an effective local temperature for one-dimensional fermionic quantum systems undergoing a Gaussian Preserving Dynamics. The identification relies on the comparison between the actual Gaussian state, generated by a dynamics which can be in general non-unitary, and a reference Gibbs state which for quadratic Hamiltonians is Gaussian too. 
The effective local temperature is deduced from the minimization of the distance between the two states, using the Hilbert-Schmidt norm of trace class operators, and leads to a  cumbersome formula which drastically simplifies in the high temperature limit. 
In this regime, the effective temperature is proportional  to the projection of the actual one particle Hamiltonian, which represents the instantaneous coupling content of the evolving system, on the reference one-particle Hamiltonian. 
Decomposing the Hamiltonian of the system on a set of local charges, representing different orthogonal energy components of the system, we show that the part of the actual coupling matrix which is orthogonal to the initial charge content does not contribute to the value of the effective inverse local temperature. 
We have applied this theory to the case of a GPD of one-dimensional hopping fermions submitted to projective measurements of the fermions densities. In such a case, we have shown numerically that the local effective temperature increases with an exponential law toward infinity as expected, see \cite{Coppola}. 
%since under the projective measurement of the local densities, the state is projected to a mixed state of 

\ukrainianpart

\title{Деякі міркування щодо локальної термалізації нерівноважних просторових квантових систем}
\author{М. Коппола, Д. Каревскі}

\address{Університет Лотарингії, CNRS, LPCT, F-54000 Нансі, Франція}

\makeukrtitle

\begin{abstract}
	Обговорюється можливість визначення локальної температури у просторових квантових системах багатьох частинок, які знаходяться далеко від стану рівноваги.	Для найпростішої модельної системи вільних ферміонів отримано явний вираз для ефективної температури у випадку, коли гаусові динамічні стани системи не завжди задовольняють умову унітарності.
	У цьому наближенні ми розглядаємо перескокові механізми міграції ферміонів на одновимірній ґратці під впливом випадкових проективних вимірювань локальних чисел заповнення. На основі усереднення за багатьма квантовими траєкторіями показано, що ефективна температура експоненційно прямує до безмежності.
	\keywords квантова статистика, нерівноважна статистична механіка, відкриті квантові системи
\end{abstract}

\lastpage

\begin{thebibliography}{10}
\bibitem{Calabrese2005}  Calabrese P.,  Cardy J. L.,  
J. Stat. Mech: Theory Exp., 2005,  P04010,
\doi{10.1088/1742-5468/2005/04/P04010}.

\bibitem{Amico} Amico L.,  Fazio R.,  Osterloh A.,  Vedral V.,
Rev. Mod. Phys., 2008, {\bf 80}, 517,
\doi{10.1103/RevModPhys.80.517}.

\bibitem{Eisler2007}  Eisler V.,  Peschel I., 
J. Stat. Mech: Theory Exp., 2007, P06005, 
\doi{10.1088/1742-5468/2007/06/P06005}.
 
\bibitem{Eisler2008}  Eisler V., Karevski D.,  Platini T.,  Peschel I., 
J. Stat. Mech: Theory Exp., 2008, P01023, 
\\\doi{10.1088/1742-5468/2008/01/P01023}.


\bibitem{Collura2010} Collura M., Karevski D.,
Phys. Rev. Lett., 2010, {\bf 104}, 200601,
\doi{10.1103/PhysRevLett.104.200601}.

\bibitem{Collura2011}  Collura M.,  Karevski D.,
Phys. Rev.   A, 2011, {\bf 83}, 023603, 
\doi{10.1103/PhysRevA.83.023603}.

\bibitem{Scopa2017} Scopa S.,  Karevski D.,
J. Phys. A: Math. Theor., 2017, {\bf 50}, 425301,
\doi{10.1088/1751-8121/aa890f}.

\bibitem{Scopa2018} Scopa S.,  Unterberger J.,  Karevski D.,
J. Phys. A: Math. Theor., 2018, {\bf 51}, 185001,
\doi{10.1088/1751-8121/aab8a5}.

\bibitem{Eisert2010}  Eisert J.,  Cramer M., Plenio M. B.,
Rev. Mod. Phys., 2010, {\bf 82}, 277,
\doi{10.1103/RevModPhys.82.277}.

\bibitem{Laflorencie}  Laflorencie N., 
Phys. Rep., 2016, {\bf 646}, 1,
\doi{10.1016/j.physrep.2016.06.008}.

\bibitem{LiebRobinson}  Lieb E. H.,  Robinson D. W.,
Nachtergaele B., Solovej  J. P., Yngvason J. (Eds.), Statistical Mechanics. Springer, Berlin, Heidelberg,  1972,
\doi{10.1007/978-3-662-10018-9_25}

\bibitem{Alba2018}  Alba V., 
Phys. Rev. B, 2018,   {\bf  97}, 245135,
\doi{10.1103/PhysRevB.97.245135}.

\bibitem{Rigol2008} Rigol M., Dunjko V.,  Olshanii M., 
Nature, 2008, {\bf 452}, 854,
\doi{10.1038/nature06838}.

\bibitem{Ilievski2015} Ilievski E.,  de Nardis J., Wouters B.,  Caux J.-S.,  Essler F. H. L.,  Prosen T.,
Phys. Rev. Lett., 2015,  {\bf 115}, 157201,
\doi{10.1103/PhysRevLett.115.157201}.

\bibitem{Ilievski2016}  Ilievski E., Medenjak M.,  Prosen T.,  Zadnik L.,
J. Stat. Mech: Theory Exp., 2016, {\bf 2016}, 064008,
\\\doi{10.1088/1742-5468/2016/06/064008}.

\bibitem{Abanin2019}Abanin  D. A., Altman E., Bloch I.,  Serbyn M.,
Rev.  Mod. Phys., 2019, {\bf  91}, 021001,
\\\doi{10.1103/RevModPhys.91.021001}.

\bibitem{Wiseman} Wiseman H. M.,  Milburn G. J., Quantum Measurement and Control, 
Cambridge University Press, Cambridge, 2009, 
\doi{10.1017/CBO9780511813948}.


\bibitem{Cao}Cao X.,  Tilloy A.,   De Luca A., 
SciPost Phys., 2019, {\bf 7}, 024,
\doi{10.21468/SciPostPhys.7.2.024}. 

\bibitem{Skinner} Skinner B.,  Ruhman J.,  Nahum A., 
Phys. Rev. X, 2019, {\bf 9}, 031009,
\doi{10.1103/PhysRevX.9.031009}.

\bibitem{Alberton} Alberton O.,  Buchhold M., Diehl S., 
Phys. Rev. Lett., 2021, {\bf  126}, 170602, 
\doi{10.1103/PhysRevLett.126.170602}.


\bibitem{Coppola}  Coppola M.,  Tirrito E.,  Karevski D.,   Collura M.,
Phys. Rev. B, 2022, {\bf 105}, 094303,
\\\doi{10.1103/PhysRevB.105.094303}.


\bibitem{Breuer} 
Breuer H.-P., Petruccione F., The Theory of Open Quantum Systems, Oxford University Press, Oxford, 2007, 
\doi{10.1093/acprof:oso/9780199213900.001.0001}.

\bibitem{Karevskibook} 
Karevski D., Physique quantique des champs et des transitions de phase, Ellipses, R\'ef\'erences sciences, 2022.

\bibitem{Karevski2009} Karevski D.,  Platini T.,
Phys. Rev. Lett., 2009, {\bf  102}, 207207,
\doi{10.1103/PhysRevLett.102.207207}.


%\bibitem{Prosen2008} Prosen T., 
%New J. Phys., 2008, {\bf  10}, 043026,
%\doi{10.1088/1367-2630/10/4/043026}


\bibitem{Platini2010}  Platini T.,  Harris R. J.,  Karevski D., 
J. Phys. A: Math. Theor., 2010, {\bf 43}, 135003,
\doi{10.1088/1751-8113/43/13/135003}.

\bibitem{Prosen2011a}  Prosen T., 
Phys. Rev. Lett., 2011, {\bf 106}, 217206,
\doi{10.1103/PhysRevLett.106.217206}.


\bibitem{Prosen2011b}   Prosen T., 
Phys. Rev. Lett., 2011, {\bf 107}, 137201,
 \doi{10.1103/PhysRevLett.107.137201}.
 
\bibitem{Karevski2013} Karevski D.,  Popkov V., Sch\"utz G. M.,
Phys. Rev. Lett., 2013, {\bf  110}, 047201,
\\\doi{10.1103/PhysRevLett.110.047201}.
 
 \bibitem{Popkov}  Popkov V.,  Karevski D.,   Sch\"utz G. M.,
Phys. Rev. E, 2013, {\bf 88}, 062118,
\doi{10.1103/PhysRevE.88.062118}.

\bibitem{Karevski2014} Landi G. T.,  Novais E.,  de Oliveira M. J.,  Karevski D., 
Phys. Rev. E, 2014, {\bf 90}, 042142,
\\\doi{10.1103/PhysRevE.90.042142}.


\bibitem{Landi2015}  Landi G. T.,  Karevski D.,
Phys. Rev. B, 2015, {\bf 91}, 174422,
\doi{10.1103/PhysRevB.91.174422}.

%\bibitem{Prosen2015} Prosen T.,
%J. Phys. A: Math. Theor., 2015, {\bf 48}, 373001,
%\doi{10.1088/1751-8113/48/37/373001}

\bibitem{Zubarev}  Zubarev D. N., Nonequilibrium Statistical Thermodynamics, 
Consultants Bureau, New York, 1974.

\bibitem{MacLennan}  MacLennan J. A., Introduction to Nonequilibrium Statistical Mechanics, 
Prentice-Hall, NJ, 1988.

\bibitem{Antal1} Antal T.,  R\'acz Z.,   Sasv\'ari L., 
Phys. Rev. Lett., 1997, {\bf 78}, 167,
\doi{10.1103/PhysRevLett.78.167}.


\bibitem{Antal2} Antal T., R\'acz Z.,  R\'akos A.,  Sch\"utz G. M.,
Phys. Rev. E, 1998, {\bf 57}, 5184,
\doi{10.1103/PhysRevE.57.5184}.

\bibitem{Antal3} Antal T., R\'acz Z.,  R\'akos A.,  Sch\"utz G. M.,
Phys. Rev. E, 1999, {\bf 59}, 4912,
\doi{10.1103/PhysRevE.59.4912}.


\bibitem{Ogata1} Ogata Y., 
Phys. Rev. E, 2002, {\bf 66}, 016135,
\doi{10.1103/PhysRevE.66.016135}.


\bibitem{Ogata2} Ogata Y., 
Phys. Rev. E, 2002, {\bf 66}, 066123,
\doi{10.1103/PhysRevE.66.066123}.

\bibitem{Aschbacher} Aschbacher W. H.,   Pillet C.-A., 
J. Stat. Phys., 2003, {\bf 112}, 1153,
\doi{10.1023/A:1024619726273}.

\bibitem{Bernard2012} Bernard D., Doyon B., 
J. Phys. A: Math. Theor., 2012, {\bf 45}, 362001,
\doi{10.1088/1751-8113/45/36/362001}.

\bibitem{Collura1} Collura M.,  Karevski D.,
Phys. Rev. B, 2014, {\bf 89}, 214308,
\doi{10.1103/PhysRevB.89.214308}.

\bibitem{Platini1} Platini T.,  Karevski D., 
J. Phys. A: Math. Theor., 2007, {\bf 40}, 1711,
\doi{10.1088/1751-8113/40/8/002}.

\bibitem{Karevski0}  Karevski D.,
Eur. Phys. J. B, 2002, {\bf 27}, 147,
\doi{10.1140/epjb/e20020139}.

\bibitem{Platini2}
Platini T.,  Karevski D.,
Eur. Phys. J. B, 2005, {\bf 48}, 225,
\doi{10.1140/epjb/e2005-00402-2}.


\bibitem{Landi1} Landi G. T.,  Karevski D.,
Phys. Rev. E, 2016, {\bf  93}, 032122, 
\doi{10.1103/PhysRevE.93.032122}.


\bibitem{Scopa0} Scopa S., Karevski D., (unpublished).


\bibitem{Hartmann1} Hartmann M., Mahler G.,  Hess O., 
Phys. Rev. Lett., 2004, {\bf 93}, 080402,
\doi{10.1103/PhysRevLett.93.080402}.

\bibitem{Hartmann2}   Hartmann M.,  Mahler G.,   Hess O., 
Phys. Rev. E, 2004, {\bf  70}, 066148,
\doi{10.1103/PhysRevE.70.066148}.

\bibitem{Hartmann3} Hartmann M.,   Mahler G., 
Europhys. Lett., 2005, {\bf 70}, 579, 
\doi{10.1209/epl/i2004-10518-5}.


\bibitem{Eisler}  Eisler V., Legeza \"O., R\'acz Z., 
J. Stat. Mech: Theory Exp., 2006, P11013,
\\\doi{10.1088/1742-5468/2006/11/P11013}.

\bibitem{Garcia} Garcia-Saez A., Ferraro A., Acin A.,
Phys. Rev. A, 2009, {\bf 79}, 052340,
\doi{10.1103/PhysRevA.79.052340}.


\bibitem{Ferraro}   Ferraro A., Garcia-Saez A.,   Acin A.,  
Europhys. Lett., 2012, {\bf 98}, 10009,
\doi{10.1209/0295-5075/98/10009}.



\bibitem{Kliesch} Kliesch M., Gogolin C., Kastoryano M. J., Riera A.,  Eisert J.,
Phys. Rev. X, 2014, {\bf 4}, 031019,
\\\doi{10.1103/PhysRevX.4.031019}.

\bibitem{DePasquale}  De Pasquale A.,  Rossini D., Fazio R., Giovannetti V., 
Nat. Commun., 2016, {\bf 7}, 12782,
\doi{10.1038/ncomms12782}.

\bibitem{Moreno2019} Moreno C., Urbina J.-D., 
Phys. Rev. E, 2019, {\bf 99}, 062135, 
\doi{10.1103/PhysRevE.99.062135}.

\bibitem{Brunner2012}  Brunner N., Linden N.,  Popescu S.,  Skrzypczyk P., Phys. Rev. E, 2012, {\bf  85}, 051117, 
\\\doi{10.1103/PhysRevE.85.051117}.

\bibitem{Morozov2018} Morozov V., Ignatyuk V., Particles, 2018, {\bf 1}, No.~1, 285--295,
\doi{10.3390/particles1010023}.

\bibitem{Latune2019} Latune C. L.,  Sinayskiy I., Petruccione F., Quantum Sci. Technol., 2019, {\bf 4}, 025005, 
\\\doi{10.1088/2058-9565/aaf5f7}.

\bibitem{Karevski1} Collura M., Aufderheide H., Roux G., Karevski D.,
Phys. Rev. A, 2012, {\bf 86}, 013615,
\\\doi{10.1103/PhysRevA.86.013615}.


\bibitem{Karevski2}  Wendenbaum P., Collura M.,  Karevski D.,
Phys. Rev. A, 2013, {\bf 87}, 023624,
\doi{10.1103/PhysRevA.87.023624}.

\bibitem{Scopa1} Scopa S.,  Karevski D.,
J. Phys. A: Math. Theor., 2017, {\bf 50}, 425301,
\doi{10.1088/1751-8121/aa890f}.


\bibitem{Scopa2} Scopa S.,  Unterberger J.,  Karevski D.,
J. Phys. A: Math. Theor., 2018, {\bf 51}, 185001,
\doi{10.1088/1751-8121/aab8a5}.

\bibitem{Riggio}  Riggio F.,  Brun Y.,  Karevski D.,  Faribault A.,  Dubail J.,
Phys. Rev. A, 2022, {\bf 106}, 053309, \\
\doi{10.1103/PhysRevA.106.053309}.



\bibitem{Popescu} Popescu S.,  Short A. J., Winter A., 
Nat. Phys., 2006, {\bf  2}, 754,
\doi{10.1038/nphys444}.

\bibitem{Goldstein} Goldstein S., Lebowitz J. L., Tumulka R.,  Zanghi N.,
Phys. Rev. Lett., 2006, {\bf 96}, 050403,
\\\doi{10.1103/PhysRevLett.96.050403}.

\bibitem{Coppola2} Coppola M., Landi G. T., Karevski D., (unpublished).
%25
\bibitem{Prosen2008} Prosen T., 
New J. Phys., 2008, {\bf  10}, 043026,
\doi{10.1088/1367-2630/10/4/043026}.
%33
\bibitem{Prosen2015} Prosen T.,
J. Phys. A: Math. Theor., 2015, {\bf 48}, 373001,
\doi{10.1088/1751-8113/48/37/373001}.

\end{thebibliography}
\end{document}